\documentclass[12pt]{article}
\begin{document}

\newcommand{\re}{\mathop{\mathrm{Re}}}

\newcommand{\be}{\begin{equation}}
\newcommand{\ee}{\end{equation}}
\newcommand{\bea}{\begin{eqnarray}}
\newcommand{\eea}{\end{eqnarray}}

\begin{center}

{\large \bf Teleparallel equivalent of general relativity: a critical review}

\vspace{2.cm}

{Janusz Garecki\footnote{E-mail address:
garecki@wmf.univ.szczecin.pl}}\\
{\it Institute of Mathematics, University of Szczecin, Wielkopolska 15,
          70-451 Szczecin, Poland}

\end{center}

\date{\today}
\vspace{0.3cm}
\begin{center}
This work is a slightly amended lecture delivered at {\bf Hypercomplex Seminar 2010} dedicated to
Professor Roman S. Ingarden on the occasion  of the Jubilee of His 90th birthday,
17-24 July 2010, at B\c edlewo,
Poland
\end{center}
\vspace{0.3cm}

\input epsf

\begin{abstract}
After reminder some facts concerning general relativity ({\bf GR})
we pass to {\it teleparallel gravity}. We are confining to the special model
of the teleparallel gravity, which is popular recently, called {\it the teleparallel
equivalent of general relativity} ({\bf TEGR}). We are finishing with conclusion
and some general remarks.
\end{abstract}




\newpage

\section{Introduction and standard formulation of {\bf GR}}

As it is known {\bf GR} is a modern  geometrical theory of
gravity which simultaneously gives a mathematical model of the
physical spacetime.

The mathematical model of the physical spacetime in {\bf
GR} is given by a pseudo-Riemannian differential manifold (Haussdorff,
paracompact, connected, inextensible, orientable) $(M_4, g_L)$.
Here $g_L$ means a Lorentzian metric which satisfies {\it Einstein
equations}
\begin{equation}
G_\mu^{~~\nu} = {8\pi G\over c^4} T_\mu^{~~\nu}
\end{equation}
$(\alpha,~\beta,~\gamma,...,\mu,~\nu, ..., = 0,1,2,3)$\footnote{
We will identify geometrical objects with the sets of their components. Greek
indices mean {\it coordinate components} of the geometrical
objects.}

So, $g_L$, is a {\it dynamical object}.

Here $G_{\mu}^{~\nu}$ is the so-called {\it Einstein tensor}, $T_{\mu}^{~\nu}$
is the {\it matter energy-momentum tensor} (the source of the gravitational field), $c$
is the velocity of light in vacuum, and $G$ means Newtonian
gravitational constant.

The mathematical model of the physical spacetime in {\bf GR} originated from
{\it Einstein Equivalence Principle} ({\bf EEP})\cite{Will93}. The main
ingredient of this Principle is {\it universality of the free
falls} of the test bodies in a given gravitational field.

{\bf GR} reduces the gravitational interactions to some geometric
aspects of the spacetime. Namely, we have:
\begin{enumerate}
\item $g_L$ = gravitational potentials,
\item $\bigl\{^{\alpha}_{\beta ~\gamma}\bigr\}$ = gravitational
strengths,
and
\item $R^{\alpha}_{~\beta\gamma\delta}(\bigl\{\bigr\})$ =
strengths of the gravitational tidal forces.
\end{enumerate}

The symmetry group of the {\bf GR} is the infinite group $\bf {Diff
M_4}$.

The Levi-Civita connection $\bigl\{^{\alpha}_{\beta~\gamma}\bigr\}$ is
symmetric, metric and torsion-free.

Usually one uses in {\bf GR} a maximal atlas of the local charts
(local maps, coordinate patches) and {\it implicite} coordinate
frames (natural frames, holonomic frames) and coframes $\bigl(\{\partial_{\mu}\},~\{dx^{\alpha}\}\bigr)$
and coordinate components of the geometrical objects.

Every coordinate transformation
\begin{equation}
x^{\alpha^{\prime}}=  x^{\alpha^{\prime}}(x^{\beta}),~
det\bigl[{\partial x^{\alpha^{\prime}}\over\partial
x^{\beta}}\bigr]\not= 0
\end{equation}
changes coordinate frames and coframes, and coordinate components
of the geometrical objects in standard way.

In the introductory relativity textbooks \cite{Schutz} one usually says about coordinate
transformations and about transformations of the coordinate
components of the geometrical objects. In fact, it is sufficient.
Also some conservative specialist on tensor analysis follow this way \cite{Schouten}.
But one can use in {\bf GR} (and in tensor calculus also)
arbitrary frames, especially {\it non-holonomic} (or anholonomic)
frames and coframes $(\{h_a^{~\mu}~(x)\},~\{h^b_{~\alpha}(x)\}):~h_a^{~\mu}(x)h^b_{~\mu}(x) =
\delta^b_a$, $(a,b,c,d,..., = 0,1,2,3)$.
Latin indices (= anholonomic indices) numerate vectors and
covectors.

The anholonomic frames and coframes {\it are not connected  with
local coordinates}, e.g., they are neutral under coordinate
transformations. Instead of we have
\begin{equation}
\partial_{\alpha} = h^b_{~\alpha}(x)\partial_b, ~~dx^{\alpha} =
h_a^{~\alpha}(x) dx^a,
\end{equation}
or, equivalently,
\begin{equation}
{\vec e}_a:=\partial_a = h_a^{~\beta}(x)\partial_{\beta},
~~\vartheta^b := dx^b = h^b_{~\mu}(x) dx^{\mu}.
\end{equation}

Here $(x):= \{x^{\alpha}\}$ are {\it spacetime coordinates}, and $\{x^a\}$
mean {\it tangent space coordinates}.
\footnote{In {\bf GR} every tangent space is endowed with Minkowski structure.}

For coordinate frames and coframes one has
\begin{equation}
{\vec e}_a = \delta^{\beta}_a\partial_{\beta},~~\vartheta^b =
\delta^b_{\mu}dx^{\mu}.
\end{equation}

Some remarks are in order:
\begin{enumerate}
\item $\{{\vec e}_a(x)\}\equiv \{\partial_a(x)\}$ is a coordinate frame in tangent
space $T_x(M_4, g_L)$, and $\{\vartheta^b\} \equiv \{dx^b\}$ is a coordinate coframe
in the dual space space $T_x^{\ast}(M_4, g_L)$.

Differential forms $\vartheta^b = dx^b = h^b_{~\mu}(x) dx^{\mu}$
{\it are not integrable} for anholonomic frames $\{h^b_{~\mu}(x)\}:d\vartheta^b\not=0$.
\item Henceforth we will consequently use an old tensorial terminology  of J.A. Schouten,
and S. Go\l \c ab, i.e., we will call $\{h_a^{~\beta}(x)\}$ ``frame'' instead of $\{{\vec
e}_a(x)\}$, and $\{h^b_{~ \mu}(x)\}$ ``coframe'' instead of
$\{\vartheta^b\}$. It will useful in passing to teleparallel
gravity because majority of the authors working in this field uses
this terminology.
\item We permanently use standard Einstein summation convention.
\end{enumerate}

As we see, anholonomic frames and coframes in our terminology
{\it connect} the partial derivatives $\partial_{\alpha}$ and $\partial_b$,
and differentials $dx^{\alpha}$ with $dx^a$.
They also connect anholonomic components of the geometrical objects (denoted by Latin indices)
with their coordinate components (denoted by Greek indices).
Namely, one has (coordinates $\{x^{\mu}\}$ are fixed)
for a tensor field of the type (r,s)
\begin{equation}
T^{a_1...a_r}_{~~~~~~b_1...b_s}(x) = h^{a_1}_{~~\mu_1}(x)
...h^{a_r}_{~~\mu_r}(x)h_{b_1}^{~~\nu_1}(x) ...
h_{b_s}^{~~\nu_s}(x) T^{\mu_1 ... \mu_r}_{~~~~~~\nu_1 ...
\nu_s}(x),
\end{equation}
and, conversely
\begin{equation}
T^{\mu_1 ... \mu_r}_{~~~~~~\nu_1 ... \nu_s}(x)
=h_{a_1}^{~~\mu_1}(x) ... h_{a_r}^{~~\mu_r}(x)h^{b_1}_{~~\nu_1}(x)
... h^{b_s}_{~~\nu_s}(x) T^{a_1 ... a_r}_{~~~~~~b_1 ... b_s}(x).
\end{equation}

For a linear and metric connection $\omega$ one
obtains\footnote{From here we confine to anholonomic tetrads and
cotetrads (See below).}
\begin{equation}
\omega^a_{~b c}(x) = h_c^{~\nu}(x)\omega^a_{~b\nu}(x),
\end{equation}
where
\begin{equation}
\omega^a_{~b\nu}(x) = h^a_{~\lambda}(x)
\Gamma^{\lambda}_{~~\mu\nu}(x)h_b^{~\mu}(x) +
h^a_{~\rho}(x)\partial_{\nu}h_b^{~\rho}(x)
\end{equation}
is so-called {\it spin connection}.
Conversely, we  have
\begin{equation}
\Gamma^{\rho}_{~~\mu\nu}(x) =
h_a^{~\rho}(x)h^b_{~\mu}(x)\omega^a_{~b\nu}(x) +
h_a^{~\rho}(x)\partial_{\nu} h^a_{~\mu}(x).
\end{equation}

In {\bf GR} one usually uses the anholonomic frames $\{h_a^{~\mu}(x)\}$
and dual coframes $\{h^b_{~\mu}(x)\}$ which form the so-called
{\it orthonormal tetrad  and cotetrad} fields. These fields are defined
as follows
\begin{equation}
h^a_{~\mu}(x) h^b_{~\nu}(x) \eta_{ab} = g_{\mu\nu}(x),
\end{equation}
or, equivalently
\begin{equation}
h_a^{~\mu}(x) h_b^{~\nu}(x) g_{\mu\nu}(x)= \eta_{ab}.
\end{equation}
Here $\eta_{ab} = diag(1,-1,-1,-1)$ is the Minkowski metric of the
tangent spaces $T_x(M_4, g_L)$ and $g_{\mu\nu}(x)$ means the
spacetime metric $g_L$.

The transformations of the spacetimes coordinates act only on
spacetime indices (Greek indices) in standard way, whereas on the
tangent space indices (Latin indices) act only {\it local
 or global Lorentz transformations}, e.g.,
\begin{equation}
h^{\prime a}_{~~\mu} = \Lambda^a_{~b}(x) h^b_{~\mu}(x),
\end{equation}
where
\begin{equation}
\Lambda^a_{~b}(x)\eta_{ac} \Lambda^c_{~d}(x) = \eta_{bd}.
\end{equation}

For a global Lorentz transformation one has $\Lambda^a_{~b} = const.$

Tetrads are not uniquely determined by the given spacetime metric $g_{\mu\nu}(x)$
but only up to local Lorentz transformations, i.e., up to six arbitrary functions.
It is because a metric has only ten independent components  and a tetrad field
has sixteen independent components. So, for a given metric $g_{\mu\nu}(x)$
there exists $\infty^6$different classes of tetrad fields $\{h_a^{~\mu}(x)\}$
which satisfy (11)-(12)\footnote{ One class of the tetrad $[\bigl\{h_a^{~\mu} (x)\bigr\}]$
means these tetrads which are connected by a global Lorentz
transformation.}.

Contrary, given tetrad field $\{h_a^{~\mu}(x)\}$ {\it determines
unique metric}
\begin{equation}
g_{\mu\nu}(x) = h^a_{~\mu}(x) h^b_{~\nu}(x) \eta_{ab},
\end{equation}
where
\begin{equation}
h^a_{~\mu}(x) h_b^{~\mu}(x) = \delta^a_b.
\end{equation}

In {\bf GR} fundamental role plays the spacetime metric $g_{\mu\nu}(x)$ (it
is an observable), whereas the orthonormal tetrads (they are not observables) play
only an auxiliary role: they simplify calculations and they enable
us to introduce spinors into spacetime structure.

The physical foundations and standard formulation of the {\bf GR}
\footnote{We mean here {\bf EEP}, Einstein equations and mathematical
model $(M_4,g_L)$ of the physical spacetime.} have very good
observational evidence. Observational consequences of the Einstein
equations were confirmed up to $0,003\%$  in Solar System (weak gravitational field),
and up to $0,05\% $ in binary pulsars (strong gravitational
field). Universality of the free falls was confirmed up to $10^{-14}$
and some other consequences of the {\bf EEP} were confirmed up to
$10^{-23}$ (See, e.g., \cite{Will93}.).

So, up to now, we {\it needn't modify or generalize} {\bf GR}.
(Ockham razor).

We would like to emphasize that {\it we have no free parameter} in {\bf
GR}.

Fascinating is that despite this the theory has passed all the stringent tests
with favour. \footnote{In the proposed {\it generalized gravity theories} one has many
free parameters, e.g., one has 28 free parameters in metric-affine
gravity. These parameters can be adjusted in order to have agreement with
experience.}

\section{Teleparallel gravity}

This is a gravity with {\it an absolute parallelism}, i.e., with
{\it curve independent parallelism} of distant vectors and
tensors.

In this old approach (since 1928; renewed recently) the
mathematical model of the physical spacetime is based on {\it
Weitzenb\"ock  geometry} (= teleparallel geometry or geometry
with absolute parallelism).

The geometry of such a kind {\it is uniquely determined} by the given
tetrad field $\{h_a^{~\mu}\}(x)$. Namely, one has (Coordinates $\{x^{\alpha}\}$
are fixed):
\begin{enumerate}
\item Metric $g_{\mu\nu}(x):= h^a_{~\mu}(x) h^b_{~\nu}(x)
\eta_{ab}.$
\item Teleparellel connection (Weitzenb\"ock's connection)
$\Gamma^{\rho}_{~~\mu\nu}:= h_a^{~\rho}(x)\partial_{\nu}h^a_{~\mu}(x).$
\end{enumerate}
Here $h_a^{~\mu}(x) h^b_{~\mu}(x) = \delta^b_a$.

The teleparallel Weitzenb\"ock connection \footnote{In the following we will call
it ``Weitzenb\"ock connection''.} has non-vanishing
torsion $T^{\rho}_{~~\mu\nu}:= \Gamma^{\rho}_{~\nu\mu} - \Gamma^{\rho}_{~\mu\nu}$
iff the tetrads $\{h_a^{~\mu}(x)\}$ are anholonomic, and it has identically vanishing
curvature $R^{\rho}_{~\theta\mu\nu}(\Gamma)$, where
\begin{equation}
R^{\rho}_{~\theta\mu\nu}(\Gamma) :=
\partial_{\mu}\Gamma^{\rho}_{~\theta\nu} -
\partial_{\nu}\Gamma^{\rho}_{~\theta\mu} +
\Gamma^{\rho}_{~\sigma\mu}{}\Gamma^{\sigma}_{~\theta\nu} -
\Gamma^{\rho}_{~\sigma\nu}{}\Gamma^{\sigma}_{~\theta\mu}.
\end{equation}

Important remarks are in order:
\begin{enumerate}
\item Weitzenb\"ock connection {\it is metric}, i.e.,
\begin{equation}
\nabla_{\rho} g_{\mu\nu} := \partial_{\rho}g_{\mu\nu} -
\Gamma^{\alpha}_{~\mu\rho} g_{\alpha\nu} -
\Gamma^{\alpha}_{~\nu\rho} g_{\mu\alpha} \equiv 0. \footnote{But the other possible
covariant derivative
\begin{equation}
{\tilde \nabla} g_{\mu\nu}(x) := \partial_{\rho} g_{\mu\nu} -
\Gamma^{\alpha}_{~\rho\mu} g_{\alpha\nu} -
\Gamma^{\alpha}_{~\rho\nu} g_{\mu\alpha},
\end{equation}
{\it is different from zero} because Weitzenb\"ock connection
is not symmetric.}
\end{equation}
\item Torsion of the Weitzenb\"ock connection {\it is entirely
determined} by the Schouten-Van Danzig {\it anholonomy object} $\Omega^a_{~bc}(x)$,
where
\begin{equation}
\Omega^a_{~bc}(x) := h_b^{~\beta}(x)h_c^{~\gamma}(x)
\bigl[\partial_{\gamma}h^a_{~\beta}(x)-\partial_{\beta}
h^a_{~\gamma}(x)\bigr].\footnote{The anholonomity object
measures anholonomy of the used tetrad field: for a holonomic tetrads $\{h_a^{~\mu}(x)\}$
one has $\Omega^a_{~bc}(x)\equiv 0$.}
\end{equation}
Namely, we have
\begin{equation}
T^{\rho}_{~\mu\nu}(x) = h_a^{~\rho}(x) h^b_{~\mu}(x) h^c_{~\nu}(x)
\Omega^a_{~bc}(x).
\end{equation}
\item One has the following relation between the components of the
Weitzenb\"ock connection $\Gamma^{\rho}_{~\mu\nu}(x)$ and between
the components $\{^{\rho}_{~\mu\nu}\}(x)$ of the Levi-Civita
connection for the metric $g_{\mu\nu}(x)$
\begin{equation}
\Gamma^{\rho}_{~\mu\nu}(x) = \{^{\rho}_{~\mu\nu}\}(x) +
K^{\rho}_{~\mu\nu}(x),
\end{equation}
where
\begin{equation}
K^{\rho}_{~\mu\nu}(x) :={1\over 2}\bigl(T_{\mu}^{~\rho}{}_{\nu} +
T_{\nu}^{~\rho}{}_{\mu} - T^{\rho}_{~\mu\nu}\bigr)
\end{equation}
is the {\it contortion tensor}.
\item For Weitzenb\"ock connection $\Gamma^{\rho}_{~\mu\nu}(x)$
\begin{equation}
\omega^a_{~b\nu}(x)\equiv 0 \Rightarrow \omega^a_{~bc}\equiv 0,
\end{equation}
i.e., this {\it connection identically vanishes}
in the tetrads $\{h_a^{~\mu}(x)\}$ which have determined it.
\end{enumerate}

Greek, i.e., holonomic indices are raised and lowered with the
spacetime metric $g_{\mu\nu}$ and the Latin, i.e., anholonomic
indices, are raised and lowered with the Minkowski metric
$\eta_{ab}$.

The class of the tetrads $\bigl[\{h_a^{~\mu}(x)\}\bigr]$ connected
by global Lorentz transformations with $\Lambda^a_{~b} = const$
determines the same Weitzenb\"ock connection and geometry. On the
other hand, the any two tetrad fields $\{h^{\prime a}_{~~\mu}(x)\}, ~~\{h^a_{~\mu}(x)\}$
which are connected by a local Lorentz transformation
\begin{equation}
h^{\prime a}_{~~\mu}(x) =\Lambda^a_{~b}(x) h^b_{~\mu}(x)
\end{equation}
{\it determine two different Weitzenb\"ock  connections}, ${\bar\Gamma}^{\rho}_{~\mu\nu}(x)$
and $\Gamma^{\rho}_{~\mu\nu}(x)$ and {\it two different
Weitzenb\"ock geometries}.

So, the set of the all tetrads $\bigl(\{h_a^{~\mu}(x)\}\bigr)$
splits onto {\it disjoint classes} ($\infty^6$ classes\footnote{$\infty^6$
classes because the local Lorentz transformations depend on six arbitrary functions.})
which determine {\it different Weitzenb\"ock connections and geometries}.

In consequence, the symmetry group of a teleparallel gravity
consists of the group ${\bf Diff M_4}$ and the global Lorentz
group.

In the following we will confine to the very special case of the
teleparallel gravity, namely we will confine to the so-called {\it
teleparallel equivalent of general relativity} ({\bf TEGR}).

The {\bf TEGR} is a recent approach to teleparallel gravity which
is mainly developed by mathematicians and physicists from Brasil
(See, e.g., \cite{Pereira}.).

One can look on {\bf TEGR} as a new trial to rescue torsion in
theory of gravity because, up to now, {\it no experiment
confirmed the Riemann-Cartan torsion}\footnote{The Riemann-Cartan torsion
is the torsion in the Riemann-Cartan geometry.
This generalized metric geometry endowed with curvature and torsion was proposed by many
authors since 1970 \cite{TrH} as a geometric model of the physical spacetime.
In our opinion lack of experimental evidence, many ambiguities to whose torsion
leads, topological triviality of torsion and {\it Ockham razor} rather disqualify
this model \cite{Gar}.}.

The details of the standard approach to {\bf TEGR} read.

One starts with the given metric $g_{\mu\nu}(x)$. This metric determines
(up to local Lorentz transformations) the anholonomic tetrad $\{h_a^{~\mu}(x)\}$
and dual cotetrad $\{h^a_{~\mu}(x)\}$ fields, which satisfy
\begin{equation}
h^a_{~\mu}(x) h^b_{~\nu}(x) \eta_{ab} = g_{\mu\nu}(x),
\end{equation}
\begin{equation}
h^a_{~\mu}(x) h_b^{~\mu}(x) = \delta^a_b.
\end{equation}
Then, these fields determine the Weitzenb\"ock connection
\begin{equation}
\Gamma^{\rho}_{~\mu\nu}(x) = h_a^{~\rho}(x) \partial_{\nu}
h^a_{~\mu}(x),
\end{equation}
which satisfies
\begin{equation}
\bigl\{^{\rho}_{~\mu\nu}\bigr\}(x) = \Gamma^{\rho}_{~\mu\nu}(x) -
K^{\rho}_{~\mu\nu}(x).
\end{equation}

Here $\bigl\{^{\rho}_{~\mu\nu}\bigr\}(x)$ is the Levi-Civita
connection for the metric $g_{\mu\nu}(x)$.

For the Weitzenb\"ock connection $\Gamma^{\rho}_{~\mu\nu}(x)$ one
has
\begin{equation}
R^{\rho}_{~\theta\mu\nu}(\Gamma)\equiv
R^{\rho}_{~\theta\mu\nu}(\{\}) + Q^{\rho}_{~\theta\mu\nu}\equiv
0.
\end{equation}
Here
\begin{equation}
R^{\rho}_{~\theta\mu\nu}(\Gamma) := \partial_{\mu}
\Gamma^{\rho}_{~\theta\nu} - \partial_{\nu}
\Gamma^{\rho}_{~\theta\mu} +
\Gamma^{\rho}_{~\sigma\mu}{}\Gamma^{\sigma}_{~\theta\nu} -
\Gamma^{\rho}_{~\sigma\nu}{}\Gamma^{\sigma}_{~\theta\mu},
\end{equation}
\begin{equation}
R^{\rho}_{~\theta\mu\nu}(\{\}) :=
\partial_{\mu}\{^{\rho}_{~\theta\nu}\}
-\partial_{\nu}\{^{\rho}_{~\theta\mu}\}
+\{^{\rho}_{~\sigma\mu}\}{}\{^{\sigma}_{~\theta\nu}\} -
\{^{\rho}_{~\sigma\nu}\}{}\{^{\sigma}_{~\theta\mu}\},
\end{equation}
and
\begin{equation}
Q^{\rho}_{~\theta\mu\nu}:= D_{\mu}K^{\rho}_{~\theta\nu} - D_{\nu}
K^{\rho}_{~\theta\\mu} + K^{\rho}_{~\sigma\mu}{}
K^{\sigma}_{~\theta\nu} -
K^{\rho}_{~\sigma\nu}{}K^{\sigma}_{~\theta\mu}.
\end{equation}
$D_{\mu}$  is the Levi-Civita covariant derivative expressed in
terms of the Weitzenb\"ock connection, i.e.,
\begin{equation}
D_{\rho} v^{\mu} := \partial_{\rho} v^{\mu} +
\bigl(\Gamma^{\mu}_{~\lambda\rho} - K^{\mu}_{~\lambda\rho}\bigr)
v^{\lambda}.
\end{equation}

$R^{\rho}_{~\theta\mu\nu}(\Gamma)$ is the {\it main}\footnote{
Main curvature tensor because one can consider other curvatures in
Weitzenb\"ock geometry, e.g., Riemannian curvature \cite{Wan}.}
curvature tensor of the Weitzenb\"ock  geometry.

The Authors which work on {\bf TEGR}, by use the fundamental
formulas (26),(29),(30) of the Weitzenb\"ock geometry, {\it
rephrase}, step by step, all the formalism of the purely metric {\bf
GR} in terms of the Weitzenb\"ock connection $\Gamma^{\rho}_{~\mu\nu}(x)$
and its torsion $T^{\rho}_{~\mu\nu}(x)$ (Mainly in terms of
torsion).

For example:
\begin{enumerate}
\item The Einstein Lagrangian for {\bf GR}
\begin{equation}
L_E = (-) \alpha\sqrt{\vert g\vert}R(\{\}) +\partial_{\mu}
w^{\mu},
\end{equation}
where $g := det\bigl[g_{\mu\nu}\bigr]$, and
\begin{equation}
w^{\mu}:= \alpha\sqrt{\vert
g\vert}\bigl(g^{\alpha\beta}\bigl\{^{\mu}_{~
\alpha\beta}\bigr\}+
g^{\alpha\mu}\bigl\{^{\gamma}_{~\alpha\gamma}\bigr\}\bigr)
\end{equation}
is rephrased to the form
\begin{equation}
\alpha h S^{\rho\mu\nu}{}T_{\rho\mu\nu} =: L_{TEGR},
\end{equation}
where $h = det\bigl[h^a_{~\mu}\bigr]= \sqrt{\vert g\vert}$,\footnote{One obtains
in fact $\infty^6$ different $L_{TEGR}$ because $L_{TEGR}$, like $L_E$ is invariant only under
global Lorentz group. Despite that the field equations (39)-(40) are locally Lorentz invariant.
We could get localy Lorentz invariant $L_{TEGR}$ if we
rephrased $L = (-)\alpha\sqrt{\vert g\vert}R(\{\}) $.}
and
\begin{equation}
S^{\rho\mu\nu} = (-) S^{\rho\nu\mu} :={1\over
2}\bigl[K^{\mu\nu\rho}  - g^{\rho\nu}{}T^{\alpha\mu}_{~~~\alpha} +
g^{\rho\mu}{}T^{\alpha\nu}_{~~~\alpha}\bigr].
\end{equation}
\item The vacuum Einstein equations
\begin{equation}
\bigl[R_{\lambda}^{~\rho}(\{\})- {1\over
2}\delta_{\lambda}^{~\rho}R(\{\})\bigr]\sqrt{\vert g\vert} = 0
\end{equation}
are rephrased to the form
\begin{equation}
\partial_{\sigma}\bigl(hS_{\lambda}^{~\sigma\rho}) -
4{\alpha}^{(-1)}\bigl(h t_{\lambda}^{~\rho}\bigr) = 0,
\end{equation}
where
\begin{equation}
t_{\lambda}^{~\rho}= h^a_{~\lambda}J_a^{~\rho} + 4\alpha
\Gamma^{\mu}_{~\lambda\nu}{}S_{\mu}^{~\nu\rho},
\end{equation}
and
\begin{equation}
J_a^{~\rho} = (-)4\alpha
h_a^{~\lambda}{}S_{\mu}^{~\nu\rho}{}T^{\mu}_{~\nu\lambda}+4\alpha
   h_a^{~\rho}{}S^{\alpha\beta\gamma}{}T_{\alpha\beta\gamma},
\end{equation}
and so on.

$\alpha:={c^4\over 16\pi G}$.
\end{enumerate}

Then, these authors call the obtained formal reformulation of {\bf GR} in
terms of the Weitzenb\"ock geometry {\it the teleparallel equivalent of the
general relativity} ({\bf TEGR})  and conclude:
``Gravitational interaction can be described {\it alternatively}
in terms of curvature, as it is usually done in {\bf GR}, or in
terms of torsion, in which case we have the so-called teleparallel
gravity. {\it Whether gravitation requires a curved or torsional
spacetime, therefore, turns out to be a matter of convention}''.
They also assert that {\bf TEGR} ``is better than the original
{\bf GR}'' because, e.g., ``in {\bf TEGR} one can separate
gravity from inertia (on the connection level) and this
separation reads''
\begin{equation}
\bigl\{^{\alpha}_{~\beta\gamma}\bigr\} =
\Gamma^{\alpha}_{~\beta\gamma} - K^{\alpha}_{~\beta\gamma}.
\end{equation}

Following the authors which work on {\bf TEGR}, the left hand side term of the above
``separation formula'', $(\{^{\alpha}_{~\beta\gamma}\})$, represents gravity and inertia
and the right hand side terms describe inertia, $(\Gamma^{\alpha}_{~\beta\gamma})$,
and gravitation, $(K^{\alpha}_{~\beta\gamma})$, respectively.

Of course, such separation {\it contradicts} {\bf EEP} and {\it is
impossible} in standard formulation of the {\bf GR}.

We {\it cannot agree with such statements}. In our opinion, the
``teleparallel equivalent of {\bf GR}'' (What kind of
equivalence?) is only {\it formal and geometrically trivial}, non-unique
(See below) rephrase of {\bf GR} in terms of the Weitzenb\"ock
geometry. Such rephrase is, of course, {\it always possible}  not
only with {\bf GR} but also with any other purely metric theory of
gravity.

In our opinion, {\it we have no profound physical motivation} for
expression of the gravitational interaction in terms of the
teleparallel torsion because the Weitzenb\"ock torsion {\it is entirely
expressed in terms of the Van Danzig and Schouten anholonomity
object} $\Omega^a_{~bc}(x)$. So, the torsion of the teleparallel
Weitzenb\"ock connection {\it describes only anholonomy} of the used
tetrad field and, therefore, {\it it is not connected neither with the real
geometry of the physical spacetime nor with real gravity}, e.g., one can introduce Weitzenb\"ock
torsion already in flat Minkowski spacetime.

Weitzenb\"ock torsion could only describe the inertial forces in the framework of the
special relativity \footnote{In special relativity anholonomic tetrads really represent
non-inertial frames.}.

Contrary, the Levi-Civita part of the Weitzenb\"ock connection, {\it as
independent of tetrads}\footnote{The Levi Civita connection depends only on metric. It is
independent of the tetrads which determine the same spacetime metric.}
, can have {\it and surely has} the physical and geometrical meaning.

Further ctitical remarks on {\bf TEGR}.
\begin{enumerate}
\item
 {\bf TEGR} is nothing new. In fact, it is exactly the old
tetrad formulation of {\bf GR} given in the very distant past by
C. M\o ller \cite{Mol} but expressed in terms of
anholonomy of the tetrads instead of in terms of tetrads exclusively ( As
it was in M\o ller papers). For example, despite that the {\bf
TEGR} field equations are expressed in terms of torsion of the
Weitzenb\"ock geometry, they form  the system of the 10 partial
differential equations of the $2^{nd}$ order on 16 tetrads
components, like the 10 field equations of the M\o  ller's tetrad
formulation of {\bf GR}. Solving
the {\bf TEGR} equations in vacuum (or in matter) we are looking for the tetrad
components $\bigl\{h_a^{~\mu}(x)\bigr\}$ for {\it apriori given general form of the metric}
$g_{\mu\nu}(x)$; not for the components of torsion. Weitzenb\"ock connection and its torsion are
calculated later \cite{Zet}.

Therefore, the notation of the Lagrangian and the field equations
of {\bf TEGR} in terms of Weitzenb\"ock torsion {\it is only a
camouflage}: {\bf TEGR} is simply the M\o ller's tetrad
formulation of {\bf GR}, and, like M\o ller's formulation of {\bf
GR}, determines uniquely the metric only.

We would like to emphasize that one can find all the results of
the {\bf TEGR} including the {\bf TEGR} energy-momentum tensor for pure gravity
in the old M\o ller's papers.\footnote{This
`tensor'' is one of the most important results obtained in the
framework of {\bf TEGR}.}
\item
{\bf TEGR} {\it is not unique}. This follows from the fact:
given metric, $g_{\mu\nu}(x)$ has 10 intrinsic components and determines
only 10 components of the tetrads field $\bigl\{h_a^{~\mu}(x)\bigr\}$
which has 16 intrinsic components. It is a consequence of the
known fact that a given metric {\it determines tetrad field up to
local Lorentz transformations}, which form the local, six-parameters, ortochronous Lorentz
group $L_+^{\uparrow}$ defined as follows
\begin{eqnarray}
L_+^{\uparrow}&=& \bigl\{\Lambda^a_{~b}(x): \Lambda^a_{~b}(x)
\eta_{ac}{}\Lambda^c_{~d}(x) = \eta_{bd},\nonumber\\
det\bigl[\Lambda^a_{~b}(x)\bigr] &=& 1, ~~\Lambda^0_{~0} \geq
1\bigr\}.
\end{eqnarray}

The ten field equations of {\bf GR} (or {\bf TEGR}) determine the metric and also determine
only ten components of the tetrad field. The remaining
six components are lefting arbitrary functions of the spacetime
coordinates $\bigl\{x^{\alpha}\bigr\}$ and can be arbitrarily
established. It is a consequence of the local Lorentz invariance
of the {\bf TEGR}  and {\bf GR} field equations.

So, for the given metric, $g_{\mu\nu}(x)$, ({\bf GR}) there exist $\infty^6$
different classes of tetrad fields ({\bf TEGR}) and, in consequence, $\infty^6$, different
Weitzenb\"ock connections $\Gamma^{\rho}_{~\mu\nu}(x)$ (and geometries). Each of
these connections satisfies the equations
\begin{equation}
\bigl\{^{\rho}_{~\mu\nu}\bigr\}(x) = \Gamma^{\rho}_{~\mu\nu}(x) -
K^{\rho}_{~\mu\nu}(x).
\end{equation}

In the above equations the left hand side {\it is independent of
tetrads}; it depends only on metric $g_{\mu\nu}(x)$, whereas the both terms on
the right hand side {\it depend on the
class of the tetrads}\footnote{ (One) class of tetrads := the set of tetrads
$\bigl[ \bigl\{h_a^{~\mu}(x)\bigr\}\bigr]$ which are connected
by global Lorentz transformations. Class of tetrads determines
the same Weitzenb\"ock connection and geometry. Different classes
of tetrads are connected by local Lorentz transformations and
determine different Weitzenb\"ock connections and geometries.}.

As a result we obtain $\infty^6$  different Lagrangians (37) for {\bf TEGR}
and $\infty^6$ different {\bf TEGR}. This fact was already known C. M\o ller in context
of his tetrad formulation of {\bf GR}. Namely, M\o ller, in fact, also has obtained $\infty^6$
different tetrad formulations of {\bf GR} because, the 10 field equations of his tetrad
formulation of {\bf GR}, identical with Einstein equations (1),
determine the tetrad field up to local Lorentz transformations,
i.e., up to six arbitrary functions. These field equations
determine the metric only.\footnote{The same situation we have of course in the framework of the
{\bf TEGR} because the 10 field equations (40), like M\o ller's equations,
are locally Lorentz invariant.}. In order to have field equations which
would determine tetrad field completely (apart from constant
Lorentz rotations) M\o  ller has developed {\it tetrad theory of gravity} in which one has
sixteen field equations onto sixteen tetrad components.
\item
The authors which work on {\bf TEGR} assert that the formula
(43) (or (45)) gives {\it separation} of inertia $\bigl(\Gamma^{\rho}_{~\mu\nu}(x)\bigr)$
from gravity $\bigl(K^{\rho}_{~\mu\nu}(x)\bigr)$.

Such speculative separation allows them, among other things, to
introduce an energy-momentum tensor for gravity \footnote{It is
in fact a family of $\infty^6$ different tensors the same as the family of the tensors
which has been obtained many years ago by C. M\o ller {\it without any separation}
in his tetrad formulation of {\bf GR}.}. But this separation {\it
is illusoric} because there exist $\infty^6$ different
separations of the form (43) (or (45)) for given $\bigl\{^{\alpha}_{~\beta\gamma}\bigr\}$,
i.e., we {\it have no separation inertia from gravity} in {\bf TEGR}
(In agreement with {\bf EEP}).

In consequence, {\it we have no unique gravitational energy-momentum
tensor} in {\bf TEGR}.
\item
The experts on {\bf TEGR} transform trivially  the geodesic
equations of {\bf GR}
\begin{equation}
{d^2x^{\alpha}\over ds^2} + \bigl\{^{\alpha}_{~
\beta\gamma}\bigr\}{dx^{\beta}\over ds}{dx^{\gamma}\over ds} = 0
\end{equation}
onto the {\it forces equations}
\begin{equation}
{d^2x^{\alpha}\over ds^2} + \Gamma^{\alpha}_{~\beta\gamma}
{dx^{\beta}\over ds}{}{dx^{\gamma}\over ds} =
K^{\alpha}_{~\beta\gamma} {dx^{\beta}\over ds}{dx^{\gamma}\over
ds}
\end{equation}
by putting in (46)
\begin{equation}
\bigl\{^{\alpha}_{~\beta\gamma}\bigr\} =
\Gamma^{\alpha}_{~\beta\gamma} - K^{\alpha}_{~\beta\gamma}.
\end{equation}

The {\it forces equations} (47) remind the {\bf GR}
equations of motion for a charged test particle when the both
fields, electromagnetic and gravitational, simultaneously
act on the particle
\begin{equation}
{d^2x^{\alpha}\over ds^2} + \bigl\{^{\alpha}_{~\beta\gamma}\bigr\}
{dx^{\beta}\over ds} {dx^{\gamma}\over ds} = {Q\over
m}F^{\alpha}_{~\beta}{dx^{\beta}\over ds}.
\end{equation}
Here $Q ,~m$ denote electric charge and mass of the particle
respectively and $F^{\alpha}_{~\beta}$ mean electromagnetic field
acting on the particle. \footnote{The right hand side of (49) is the
electromagnetic force per unit mass which acts on the particle.}
\end{enumerate}

The specialists on {\bf TEGR} try to attach some physical meaning
to the {\it force equations} (47), namely following them, the right hand side of (47) describes
{\it gravitational force} acting on the particle, whereas the term
$\Gamma^{\alpha}_{~\beta\gamma}{dx^{\beta}\over ds}{dx^{\gamma}\over ds}$ describes
{\it inertial force}.

But there exist $\infty^6$
different reformulatios of the geodesic equations (46) to the form
(47) with different $\Gamma^{\alpha}_{~\beta\gamma}$ and
$K^{\alpha}_{~\beta\gamma}$. Which one of them is correct, i.e.,
{\it which one of them gives correct inertial force and correct
gravitational force}?

Talking  about {\it equivalence} of  {\bf TEGR} with {\bf GR} {\it is
misleading} because there exist $\infty^6$ different {\bf
TEGR} in consequence of the local Lorentz invariance  of the field equations (40)
\footnote{But we must emphasize that every {\bf TEGR}
determines unique and the same metric structure of the spacetime as {\bf GR}
does. So, from the metric point of view, the different {\bf TEGR} are equivalent.}.

Here we have the same kind of {\it ``equivalence''} as the {\it
``equivalence''} between a given metric $g_{\mu\nu}(x)$ (10 functions) and a tetrad
field (16 functions), which satisfies $h^a_{~\mu}(x)h^b_{~\nu}(x)\eta_{ab} = g_{\mu\nu}(x)$
i.e., {\it we have no equivalence}
\footnote{Remark also that metric and tetrads are different geometric
objects.}.

Incorrect is also statement of the specialists on {\bf TEGR} that
Weitzenb\"ock geometry is flat, like Minkowski geometry. In fact, e.g.,
{\it Riemannian curvature of such geometry is non-zero}. Also the curvature tensor
${\tilde R}^{\alpha}_{~\beta\gamma\delta}(\Gamma)$
where
\begin{equation}
{\tilde R}^{\alpha}_{~\beta\gamma\delta}(\Gamma):=
\partial_{\gamma} \Gamma^{\alpha}_{~\delta\beta} -
\partial_{\delta} \Gamma^{\alpha}_{~\gamma\beta} +
\Gamma^{\alpha}_{~\gamma\sigma}{}\Gamma^{\sigma}_{~\delta\beta} -
\Gamma^{\alpha}_{~\delta\sigma}{}\Gamma^{\sigma}_{~\gamma\beta}
\end{equation}
{\it is different from zero}.

The tensor ${\tilde R}^{\alpha}_{~\beta\gamma\delta}(\Gamma)$
differs from the former {\it main curvature tensor} $R^{\alpha}_{~\beta\gamma\delta}(\Gamma)$
(See the formula (31)) by transposition lower indices in
$\Gamma^{\alpha}_{~\beta\gamma}(x)$.\footnote{For Riemannian
geometry, owing to symmetry of the Levi-Civita connection, these both tensors
are identically equal.}

Resuming, in our opinion, {\bf TEGR} is nothing new. It is
camouflaged, the very old tetrad formulation of {\bf GR} given by
C. M\o ller, and it, {\it by no means is better } than standard {\bf
GR}. Contrary, standard {\bf GR} {\it is surely better} than any {\bf
TEGR} because  {\bf GR} is invariand under any change of
tetrads, whereas {\bf TEGR} is not. {\bf TEGR}, like any teleparallel gravity, is invariant only
under global Lorentz rotations of tetrads.

We will finish with some general remarks about
teleparallel gravity.

It should be emphasized that there exist many other
approaches to teleparellel gravity, different from {\bf TEGR},
and which generalize {\bf GR}. At the first time such approach to
gravity was considered already by A. Einstein
(``Fernparallelismus'' in 1928 \cite{E28}) and then by C. M\o ller (1978),
Pellegrini and Plebanski \cite{Pleb}, Hayashi and Shirafuji \cite{Hay}, and others.
Recently the teleparallel approach to gravity is developed by F.B. Estabrook,
Y. Itin, and L. Sch\"ucking \cite{Itin}.

In these other approaches to teleparallel gravity the gravitational Lagrangian is built
from irreducible torsion componets or from tetrads immediately, and contains, in general,
three free parameters to be determined by experiments. This
Lagrangian is invariant  under ${\bf Diff M_4}$ and has also
global Lorentz symmetry.

The fundamental geometric object are  tetrads which determine
spacetime metric and Weitzenb\"ock connection, and, therefore, all
the local Weitzenb\"ock geometry of the physical spacetime.

In vacuum, we have in these approaches sixteen $2^{nd}$ order field
equations on sixteen tetrad components. The field equations
should determine the tetrads field $h_a^{~\mu}(x)$ up to constant
Lorentz rotations, i.e., up to global Lorentz group, and owing that, should determine
a unique Weitzenb\"ock geometry. But tetrads {\it are not
observables}: they are very alike to the electromagnetic
potentials. Moreover, there are problems with physical interpretation
of the six additional tetrads components (10 components can
describe gravitational field, but what about remaining 6
components?) and these theories suffer from badly posed Cauchy problem
\cite{Kop}.

\begin{center}
{\bf Acknowledgments}
\end{center}

The Author would like to thank Professor Julian  \L awrynowicz for
possibility to deliver this Lecture during the {\bf Hypercomplex
Seminar 2010} dedicated to Professor Roman S. Ingarden on the occasion of His 90th
birthday, and the Mathematical Institute of the University of Szczecin for financial
support (grant 503-4000-230351). Author also thanks Professor
Friedrich W. Hehl for the most useful critical remarks.

\newpage
\begin{center}
{\bf Teleparalelny ekwiwalent og\'olnej teorii wzgl\c
edno\'sci: uwagi krytyczne}
\end{center}

\vspace{0.3cm}
\begin{center}
Janusz Garecki
\end{center}

\begin{center}
Instytut Matematyki Uniwersytetu Szczeci\'nskiego
\end{center}

\begin{center}
 Streszczenie
\end{center}

Po przedstawieniu  podstawowych fakt\'ow z og\'olnej teorii wzgl\c
edno\'sci oraz z teleparalelnej grawitacji, ograniczam si\c e  do
analizy specjalnego modelu teleparalelnej grawitacji nazwanego
przez jego tw\'orc\'ow {\it teleparalelnym ekwiwalentem og\'olnej
teorii wzgl\c edno\'sci} (w skr\'ocie {\bf TEGR}). Model ten by\l
  (i jest) ostatnio intensywnie badany g\l \.ownie przez matematyk\'ow i
 fizyk\'ow z Brazylii.

W pracy pokazuj\c e, \.ze {\bf TEGR} jest zakamuflowanym, starym,
tetradowym sformu\l owaniem og\'olnej teorii wzgl\c
edno\'sci, dokonanym w latach 60-tych i 70-tych XX-go wieku przez
C. M\o llera i podkre\'slam, \.ze {\bf TEGR} jest niejednoznacznym
i trywialnym przeformu\l owanie og\'olnej teorii wzgl\c edno\'sci,
kt\'ore nie mo\.ze da\'c nic lepszego od standardowe sformu\l
owanie tej teorii (Moim zdaniem, przeformu\l owanie to jest gorsze).


\begin{thebibliography}{99}
\bibitem{Will93} C.M. Will, ``Theory and Experiment in
Gravitational Physics'', Cambridge University Press, Cambridge
1993; ``The Confrontation between General Relativity and Experiment'', arXiv:gr-qc/0510072.
\bibitem{Schutz}B.F. Schutz, ``A First Course in General
Relativity'', Cambridge University Press, Cambridge 1985 ( Polish
edition: B.F. Schutz,~``Wst\c ep do Og\'olnej Teorii Wzgl\c edno\'sci'',
Wydawnictwo Naukowe PWN, Warszawa 2002); L.D. Landau, E.M. Lifshitz,
``The Classical Theory of Fields'', 4th edition, Oxford 2002
(Polish edition: L.D. Landau, E.M. Lifszyc, ``Teoria Pola'',
Wydawnictwo Naukowe PWN, Warszawa 2009); J. Foster, J.D.
Nightingale,``A Short Course in General Relativity'', Longman,
London and New York 1979 (Polish edition: J. Foster, J.D.
Nightingale, Og\'olna Teoria Wzgl\c edno\'sci'', PWN, Warszawa 1985);
J. Pleba\'nski and A. Krasi\'nski, ``An Introduction to General
Relativity and Cosmology'', Cambridge University Press, Cambridge
2006; S. Carrol, ``Spacetime and Geometry. An Introduction to
General Relativity'', Addison Wesley, 2004; W. Rindler,
``Relativity, Special, General, and Cosmological'', Oxford
University Press, Oxford 2004; A.Trautman, W. Kopczy\'nski,
``Czasoprzestrze\'n i Grawitacja'', PWN, Warszawa 1984 (There exists
English edition).
\bibitem{Schouten} J.A. Schouten, ``Ricci--Calculus'',
Springer-Verlag, Berlin 1954; S. Go\l \c ab, ``Tensor Calculus'',
PWN, Warsaw 1966 (In Polish.There exists English edition.).
\bibitem{Pereira} V.C. de Andrade, L.C.T. Guillen and J.G.
Pereira, {\it Int. J. Mod. Phys}., {\bf D13}(2004) 2193
(arXiv:gr-qc/0501017); R. Aldrovandi, T.G. Lucas and J.G. Pereira,
``Does a tensorial energy-momentum density for gravitation
exist?'', arXiv:0812.0034 [gr-qc]; R. Aldrovandi, T.G. Lucas and
J.G. Pereira, ``Inertia and gravitation in teleparallel gravity'',
arXiv:0812.0034 v.2 [gr-qc]; V.C. de Andrade, H.I. Arcos and J.G.
Perei
ra, ``Torsion as Alternative to Curvature in the Description
of Gravitation'', arXiv: gr-qc/0412034; H.I. Arcos, T.G. Lucas and
J.G. Pereira, ``A Consistent Gravitationally--Coupled Spin--2Field
Theory'', arXiv:1001.3407 [gr-qc]; R.Aldrovandi, J.G. Pereira and
K.H. Vu, ``Doing without the Equivalence principle'',
arXiv:gr-qc/0410042; R.A. Mosua and J.G. Pereira, {\it Gen. Rel.
Gravit.,} {\bf 36} (2004) 2525 (arXiv:gr-qc/0312093); R.
Aldrovandi and J.G. Pereira, ``Gravitation: in search of the
missing torsion'', arXiv:0801.4148 [gr-qc]; H.I. Arcos and J.G.
Pereira, ``Torsion and the gravitational interaction'',
arXiv:gr-qc/0408096; J.G. Pereira, T. Vargas and C.M. Zhang, ``
Axial--Vector Torsion and the Teleparallel Kerr Spacetime'',
arXiv:gr-qc/0102070; V.C. de Andrade, L.C.T. Guillen and J.G.
Pereira, ``Teleparallel gravity: an overview'',
arXiv:gr-qc/0011087; T.G. Lucas, Y.N. Obukhov and  J.G. Pereira,
`` Regularizing role of teleparallelism'', arXiv:0909:2418
[gr-qc]; H.I. Arcos, V.C. de Andrade and J.G. Pereira, `` Torsion
and Gravitation: a New View'', arXiv:gr-qc/0403074; J.W. Maluf,
F.F. Faria and K.H. Castello-Branco, {\it Class. Quantum Grav}.,
{\bf 20} (2003) 4683; V. C. de Andrade, L.C.T. Guillen and J.G.
Pereira, ``Teleparellel Spin Connection'', arXiv:gr-qc/0104103;
Y.N. Obukhov and J.G. Pereira, {\it Phys. Rev}., {\bf D67} (2003)
044008; A.A. Sousa, R.B. Pereira and A.C. Silva, ``Energy and
angular momentum densities in a G\"odel--type universe in the
teleparallel  geometry'', arXiv:0803.1481 [gr-qc]; J.F. da
Rocha-Neto and K.H. Castello-Branco, ``Gravitational Energy of
Kerr and Kerr Anti-de Sitter Space-Times in the Teleparallel
Geometry'', arXiv:gr-qc/0205028; J.W. Maluf, J.F. da
Rocha-Neto, T.M.L. Toribio and K.H. Castello-Branco, ``Energy and
angular momentum of the gravitational field in the teleparallel
geometry'', arXiv:gr-qc/0204035; M. Sharif and A.Jawal, ``Energy
Content of Some Well-Known Solutions in Teleparallel Gravity'',
arXiv:1005.5203 [gr-qc];  R. Aldrovandi, J.G. Pereira and K.H. Vu,
``Selected Topics in Teleparallel Gravity'', arXiv:gr-qc/0312008;
R. Aldrovandi, P.B. Barros and J.G. Pereira, {\it Gen. Rel.
Gravit}., {\bf 35} (2003) 991; V.C. de Andrade, L.C. T. Guillen
and J.G. Pereira, ``Teleparallel Gravity  and the Gravitational
Energy-Momentum Density'', arXiv:gr-qc/0011079; R. Aldrovandi and
J.G Pereira, ``An introduction to teleparallel gravity'',
Instituto de Fisica Teorica, UNESP, Sauo Paulo, Brasil 2007.
\bibitem{TrH} A. Trautman, ``On the Structure of the
Einstein-Cartan Equations'', Istituto Nazionale di Alta
Matematica, {\it Symposia Mathematica}, {\bf 12} (1973) 139; F.W.
Hehl, {\it Gen. Rel. Gravit}., {\bf 4} (1973) 333; F.W. Hehl, --ibidem {\bf
5} (1974) 491; F.W. Hehl et al., {\it Rev. Mod. Phys.}, {\bf 48}
(1976) 393; F.W. Hehl et al., ``Gravitation  and the Poincar\'e
Gauge Field Theory with Quadratic Lagrangian'', in {\it General
Relativity and Gravitation}, Vol.I., Ed.A. Held, Plenum Publishing
Corporation 1980; A. Trautman, ``Fiber Bundles, Gauge Fields and
Gravitation'', in {\it General Relativity and Gravitation}, Vol.I.
Ed. A. Held, Plenum Publishing Corporation 1980; A. Trautman,
``Einstein-Cartan Theory'', arXiv:gr-qc/0606062.
\bibitem{Gar} J. Garecki, ``On Torsion in a Theory of Gravity'',
in {\it Relativity, Gravitation, Cosmology}, Eds. V. Dvoeglazov
and A. Espinoza Garrido, 2004 Nova Science Publishers, Inc.
\bibitem{Wan} M.I. Wanas, ``Absolute Parallelism Geometry:
Developments, Applications and Problems'', arXiv:gr-qc/0209050.
\bibitem{Mol} C. M\o ller, ``Conservation laws in the tetrad
theory of gravitation'', in {\it Relativistic Theories of
Gravitation}, Ed. L. Infeld, Pergamon Press, Oxford$\bullet$ London$\bullet$ Edinburgh$\bullet$
New York$\bullet$ Paris$\bullet$ Frankfurt. Copyright 1964 by PWN,
Warszawa 1964; C. M\o ller, {\it Mat. Fys. Medd. Dan. Vid.
Selsk.}, {\bf 35} (1966) 14pp (NORDITA publications No.190); C.
M\o ller, ``On the Crisis in the Theory of Gravitation and a
Possible Solution'', {\it Mat. Fys. Medd. Dan. Vid. Selsk.}, {\bf
39} (1978) 31pp, K\o  benhavn 1978.
\bibitem{Zet} G. Zet, `` Schwarzschild Solution on a Space-Time
with Torsion'', arXiv:gr-qc/0308078.
\bibitem{E28} T. Sauer, `Field equations in teleparallel
spacetime:Einstein's {\it Fernparallelismus} approach towards
unified field theory'', arXiv:physics/0405142.
\bibitem{Pleb} J. Pleba\'nski,``Tetrads and Conservation Laws'',
in {\it Relativistic Theories of Gravitation}, Ed.L. Infeld, Pergamon Press,
 Oxford$\bullet$London$\bullet$Edinburgh$\bullet$New York$\bullet$Paris$\bullet$Frankfurt.
Copyright 1964 by PWN, Warszawa 1964.
\bibitem{Hay} K. Hayashi and T. Shirafuji, {\it Phys. Rev.}, {\bf
D 19} (1979) 3524; F. M\"uller-Hoissen, ``On the tetrad theory of
gravity'', MPI--PAE/Pth61/84, 17pp.
\bibitem{Itin} Y. Itin, ``Coframe geometry and gravity'',
arXiv:0711.4209 [gr-qc]; Y. Itin, ``Does the coframe geometry can
serve as a unification scheme?'', arXiv:gr-qc/0409071; E.L.
Sch\"ucking, ``Einstein's Apple and Relativity's Gravitational
Field'', arXiv:0903.3768 [physics.hist-ph]; F.B. Estabrook, ``Conservation
Laws for Vacuum Tetrad Gravity'', arXiv:gr-qc/0508081.
\bibitem{Kop} W. Kopczy\'nski, {\it J.Phys.},{\bf A 15} (1982)
493.
\end{thebibliography}
\end{document}